# The Influence of Sodium and Potassium Dynamics on Excitability, Seizures, and the Stability of Persistent States: II. Network and Glial Dynamics.

Ghanim Ullah[1], John R. Cressman Jr.[2], Ernest Barreto[2], and Steven J. Schiff[1, 3]

[1] Center for Neural Engineering, Department of Engineering Science and Mechanics, The Pennsylvania State University, University Park, PA, 16802, USA, [2] Krasnow Institute for Advanced Study, George Mason University, Fairfax, VA, 22030, USA, and [3] Departments of Neurosurgery and Physics, The Pennsylvania State University, University Park, PA, 16802, USA.

## ABSTRACT

In these companion papers, we study how the interrelated dynamics of sodium and potassium affect the excitability of neurons, the occurrence of seizures, and the stability of persistent states of activity. We seek to study these dynamics with respect to the following compartments: neurons, glia, and extracellular space. We are particularly interested in the slower time-scale dynamics that determine overall excitability, and set the stage for transient episodes of persistent oscillations, working memory, or seizures. In this second of two companion papers, we present an ionic current network model composed of populations of Hodgkin-Huxley type excitatory and inhibitory neurons embedded within extracellular space and glia, in order to investigate the role of micro-environmental ionic dynamics on the stability of persistent activity. We show that these networks reproduce seizure-like activity if glial cells fail to maintain the proper micro-environmental conditions surrounding neurons, and produce several experimentally testable predictions. Our work suggests that the stability of persistent states to perturbation is set by glial activity, and that how the response to such perturbations decays or grows may be a critical factor in a variety of disparate transient phenomena such as working memory, burst firing in neonatal brain or spinal cord, up states, seizures, and cortical oscillations.

## INTRODUCTION

Despite an extraordinary amount of interest in understanding the dynamics of seizures, we still lack a unifying dynamical definition of what a seizure is (Soltesz and Staley, 2008). The extraordinary variety of experimental preparations and human epilepsies makes the quest for unifying principles especially difficult.

In the first of these two companion papers, we examined the role of potassium dynamics on the stability of the activity of a single neuron (Cressman et al., Submitted). In this second of two papers, we apply our findings to the stability of networks of neurons. We undertake our study using the concept of “persistent states” of activity as a paradigm for our investigation. Increasing evidence points to the importance of persistent activity in the maintenance of mental state and working memory (Miller et al., 2003; Wang, 2003). The maintenance of neural activity over time following an event has been observed experimentally in monkey prefrontal cortex during short-term working memory, i.e., delay periods of tasks that require the animals to remember certain features such as target location (Funahashi et al., 1989; Fuster, 1995; Goldman-Rakic, 1995; Miller et al., 1996; Rainer et al., 1998; Romo et al., 1999). The brain requires a certain degree of instability to form such an activity balance (often likened to a Turing instability (Murray, 2003)). We here ask: what are the network properties that render such physiological persistent states stable to perturbations, and especially: what are the circumstances where such perturbations lead to seizure-like activity.

Synaptic interaction is in many systems the driver of persistent excitation. Neurons within a network may sustain activity by circulating activity among themselves, thus bypassing individual neurons' refractory period. Such persistent network activity often requires a balance between inhibitory and excitatory activity (McCormick et al., 2003; Shu et al., 2003; Van Vreeswijk and Sompolinsky, 1996). When inhibition is strong compared to excitation, the network responds to strong stimuli, but negative feedback suppresses the network such that the activity of all neurons decays with time. In intermediate conditions, where the excitation is balanced by inhibition, the network can exhibit a stable focus of activity (see for example Durstewitz et al., 2000; Vogels et al., 2005). This mode can be utilized in different kinds of information processing, for example to sustain a representation of spatial information for a period of time (Funahashi et al., 1989). When the network is dominated by excitation, the activity of all neurons can increase to a relatively high firing rate, and information processing by the network is impaired. For example, blocking $GABA_A$–receptor mediated inhibition can result in the transformation of recurrent persistent activity into intense epileptiform activity (Sanchez-Vives and McCormick, 2000).

In most models of pathological conditions such as seizures, a shift from dominant (or balanced) inhibition to dominant excitation in a neuronal network is considered to be responsible for the network transition from the pre-ictal to ictal state (Trevelyan et al., 2006). In such conditions of imbalance between inhibition and excitation, moderate perturbations can drive a neuronal network from physiological to seizure-like activity. However, the conditions under which such network transitions occur in response to perturbations are not well known.

We pose the hypothesis that how an active neuronal network handles perturbations is a dynamical signature of its functional stability. We draw this analogy from the physics of equilibrium systems (Forster, 1990). We suggest that such perturbation responses are an operational test of excitatory-inhibitory neuronal balance in neuronal networks.

We focus on the role that the extracellular space and glia may play in modulating the stability of these persistent states. Glial cells absorb excess ions, most importantly $K^+$, from the extracellular space and thus regulate excessive excitation due to decreased neuronal transmembrane $K^+$ gradients. Glia also release neurotransmitters such as glutamate, which can cause excitation in the surrounding neurons through the activation of NMDA receptors (Parpura et al., 1994; Parpura and Haydon, 2000; Tian et al., 2005). However, a characterization of how glial dysfunction affects the dynamics of neuronal network activity remains largely unstudied. For example, what would happen if the glial cells fail to adequately remove the excess $K^+$ ions from the interstitial volume or if they suddenly release excessive glutamate while the network is performing normally?

In recent work, there is gathering evidence characterizing the interplay between excitatory and inhibitory neurons during seizures (Fujiwara-Tsukamoto et al., 2004; Perez-Velazquez and Carlen, 1999). Ziburkus, et al. (2006) observed interplay between pyramidal cells (PCs) and interneurons (INs) during *in vitro* seizure-like events. In particular, PCs were seen go into a silent state when INs were burst firing, followed by burst firing in PCs when INs went to depolarization block. What might be the underlying network dynamics that can lead to such seizure patterns?

In this second of two companion papers we address these questions through mathematical modeling. We construct a physiologically-motivated model for the neuronal network that combines the Hodgkin-Huxley formalism for the neuronal ionic currents with a model for the dynamics of the extra- and intracellular $K^+$ concentration ([K]) controlled by the glial network and active ion pumps. We also include dynamic extra- and intracellular $Na^+$ concentration ([Na]). We show that this model is able to maintain localized, persistent activity when the glial network is functioning normally and excitation and inhibition are balanced. The model is then used to explore physiological conditions under which an otherwise normal network would show abnormal activity. We show that such networks reproduce seizure-like activity if glial cells fail to maintain the proper extracellular [K]. Our model builds upon the findings of a simplified single cell ionic micro-environmental model in the accompanying manuscript (Cressman at al., Submitted).

## METHODS

A schematic of our model network is shown in Figure 1.

*Membrane potential dynamics:*

We used single-compartment neurons in a network consisting of 100 PCs and 100 INs. The equations governing the membrane potential of the PCs and INs are adopted from the model in (Gutkin et al., 2001), and are as follows.

$$C\frac{dV^{e/i}}{dt} = I_{Na}^{e/i} + I_K^{e/i} + I_L^{e/i} + I_{syn}^{e/i} + I_{ext}^{e/i} + I_{rand}^{e/i}$$

$$I_{Na}^{e/i} = -g_{Na}\left[m_\infty^{e/i}(V^{e/i})\right]^3 h^{e/i}\left(V^{e/i} - V_{Na}^{e/i}\right)$$

$$I_K^{e/i} = -\left(g_K[n^{e/i}]^4 + \frac{g_{AHP}[Ca]_i^{e/i}}{1+[Ca]_i^{e/i}}\right)(V^{e/i} - V_K^{e/i}), \qquad (1)$$

where the superscript *e/i* refers to PC/IN (excitatory/inhibitory) cells in the network. Both intrinsic and synaptic ( $I_{syn}^{e/i}$) currents are incorporated into the model neurons. The gating variables for intrinsic currents are described in the companion paper (Cressman et al., submitted). The leak current used here is lumped and is given as

$$I_L^{e/i} = -g_L(V^{e/i} - V_L^{e/i}). \qquad (2)$$

The slow calcium-activated $K^+$ current (the spike-frequency adapting or afterhyperpolarization current, $g_{AHP}$) is only included in the PCs, i.e. $g_{AHP}$ = 0 mS/cm$^2$ for INs. The various parameters and variables used in this paper are defined in Table 1.

In order to take into account the Ziburkus et al. (2006) findings of PC-IN interplay during seizure-like events, we modify the synaptic current entering the *jth* PC and IN from that used in (Gutkin et al., 2001). We use

$$I_{syn}^e = -\frac{(V_j^e - V_{ee})}{N}\sum_{k=1}^N g_{jk}^{ee} s_k^e \chi_{jk}^e - \frac{(V_j^e - V_{ie})}{N}\sum_{k=1}^N g_{jk}^{ie} s_k^i \chi_{jk}^i,$$

$$I_{syn}^{i} = -\frac{(V_j^i - V_{ei})}{N}\sum_{k=1}^{N} g_{jk}^{ei} s_k^e \chi_{jk}^e - \frac{(V_j^i - V_{ii})}{N}\sum_{k=1}^{N} g_{jk}^{ii} s_k^i \chi_{jk}^i , \tag{3}$$

where $V_j^{e/i}$ is the voltage of the jth excitatory/inhibitory neuron, $s_k^{e/i}$ is the variable giving the temporal evolution of the synaptic efficacy emanating from the kth excitatory/inhibitory neuron, and $\chi_{jk}^{e/i}$ takes into account the interplay between PCs and INs described above; that is, if a cell goes into depolarization block, then the synaptic inputs from that cell to others are reduced towards zero by the factor $\chi_{jk}^{e/i}$. $V_{ee}$, $V_{ei}$, $V_{ie}$, and $V_{ii}$ are the reversal potentials for the excitatory-excitatory, excitatory-inhibitory, inhibitory-excitatory, and inhibitory-inhibitory synaptic inputs, respectively. The subscripts *ij* in all equations in this paper represent the connection from cell *i* to cell *j*. The variables used in equations (1-3) are given as

$$\tau^{e/i}\frac{ds^{e/i}}{dt} = \phi\sigma(V^{e/i})(1 - s^{e/i}) - s^{e/i}$$

$$\sigma(V^{e/i}) = 1/[1 + \exp(-(V^{e/i} + 20)/4)]$$

$$g_{jk}^{ee} = \alpha_{ee}\sqrt{\frac{100}{\pi}}\exp(-100[(j-k)/N]^2)$$

$$g_{jk}^{ie} = \alpha_{ie}\sqrt{\frac{30}{\pi}}\exp(-30[(j-k)/N]^2)$$

$$g_{jk}^{ei/ii} = \alpha_{ei/ii}\sqrt{\frac{30}{\pi}}\exp(-30[(j-k)/N]^2)$$

$$\chi_{jk}^{e/i} = \begin{cases} \exp(-\eta^{e/i}/\upsilon) & if \quad \eta^{e/i} > 5.0 \\ 1 & otherwise \end{cases}$$

$$\frac{d\eta^{e/i}}{dt} = \gamma^{e/i}(V^{e/i} - V_b) - \tilde{\gamma}\,\eta^{e/i}$$

$$\gamma^{e/i} = \begin{cases} 0.4 & if \quad -30 < V^{e/i} < -10 \\ 0 & otherwise \end{cases} . \tag{4}$$

The parameter values used in equations (1-4) are $\phi = 3$, $\tau^e = 4$ms, $\tau^i = 8$ms, $g_L = 0.05$ mS/cm$^2$, $V_{ee} = 0$mV, $V_{ie} = -80$mV, $\alpha_{ie} = 0.06$, $V_b = -50.0$mV, $\tilde{\gamma} = 0.4$ and $\upsilon = 5.0$, $V_{ei} = 0$mV, $V_{ii} = -80$mV. The factors -100, -30, and -30 inside the exponential terms in the expressions for $g_{jk}^{ee}$, $g_{jk}^{ie}$, and $g_{jk}^{ei/ii}$, respectively, model the different spatial ranges of synaptic inputs made by the PCs and INs; thus, the PC-to-PC connections have a narrow spatial footprint, while the PC-IN and the IN-IN synaptic footprints are wider. $\chi_{jk}^{e/i}$ models the synaptic block due to depolarization, $\upsilon$ is an associated scale, $V^{e/i} \in$ [-30,-10], and $\eta > 5$ is the threshold to synaptic block due to depolarization.

Random currents ( $I_{rand}^{e,i}$ ) are chosen from a uniform distribution over [-10, 10], where both negative and positive pulses have equal arrival probability. A constant external current $I_{ext}^{i}$ = 0.5 is injected into the INs. For simulations in this paper (unless otherwise mentioned), a spatially restricted stimulus of Gaussian form $I_{ext}^{e} = I_{amp} \exp(-60[(j - N/2)/N]^2)$ was injected into PCs 21-79 at t = 112ms for 20ms to induce the stable focus of activity, where $N$ = 100, $j$ is neuron number and $I_{amp}$ = 1.5 (see also Gutkin et al., 2001). A value of $\alpha_{ii}$ = 0.02 is used in all simulations.

*Ion concentrations dynamics:*

The model also contains dynamic extra- and intracellular [K] and [Na], subject to the constraints in Cressman et al. (Submitted); see also Eqns. (8) and (9) below. The [K] in the interstitial volume surrounding each cell ($[K]_o$) was continuously updated based on $K^+$ currents across the neuronal membrane, $Na^+$-$K^+$ pumps, uptake by the glial network surrounding the neurons, and lateral diffusion of $K^+$ within the extracellular space. The $[K]_o$ dynamics is modeled by the following differential equation, which is modified from that in (Cressman et al., Submitted) because of the network structure:

$$\frac{d[K]_o^{e/i}}{dt} = 0.33 I_K^{e/i} - 2\beta I_{pump}^{e/i} - I_{diff}^{e/i} - I_{glia}^{e/i} + \frac{D}{\Delta x^2}([K]_{o(+)}^{e/i} + [K]_{o(-)}^{e/i} + [K]_o^{i/e} - 3[K]_o^{e/i}). \quad (5)$$

The factor $\beta$ = 7.0 corrects for the volume fraction between the interior of the cell and the extracellular space when calculating the concentration change and is based on (Mazel et al., 1998; McBain et al., 1990; Somjen, 2004). $[K]_{o(+)}^{e/i}$ and $[K]_{o(-)}^{e/i}$ are the $K^+$ concentrations in the adjacent extracellular volumes corresponding to nearest neighbors in the same layer, $[K]_o^{i/e}$ is the $K^+$ concentration around the nearest neighbor in the second layer and mimics the $K^+$ diffusion across the layers, and $D = 2.5\times10^{-6}$cm$^2$/s is the diffusion coefficient for $K^+$ in water (Fisher et al., 1976). The factor 0.33mM.cm$^2$/μcoul in front of $I_K^{e/i}$ is used to convert the currents into ion concentration rate-of-change, and 2.0 in front of $\beta I_{pump}$ is due to the fact that the $Na^+$-$K^+$ pump has an electrogenic 2:3 ratio. Further discussion of $\beta$ and the 0.33 can be found in Appendix A from the companion paper Cressman, et al. (Submitted).

The model also includes terms that replicate the effects of active pumps and diffusion through the surrounding glial network (Cressman et al., Submitted). For pumps, we use

$$I_{pump}^{e/i} = \left(\frac{1.25}{1.0+\exp((25.0-[Na]_i^{e/i})/3.0)}\right)\left(\frac{1.0}{1.0+\exp(8.0-[K]_o^{e/i})}\right). \quad (6)$$

The $K^+$ diffusion between the extracellular space and the bath solution is modeled by

$$I_{diff}^{e/i} = \varepsilon([K]_o^{e/i} - k_{o,\infty}),$$

and the glial uptake is modeled by

$$I_{glia}^{e/i} = -\frac{G_{glia}}{1.0+\exp((18-[K]_o^{e/i})/2.5)}, \quad (7)$$

where $k_{o,\infty}$ is the steady state $K^+$ concentration in the nearby reservoir – either the bath solution in a slice preparation, or the vasculature in the intact brain. Unless otherwise mentioned, a value of 1.3sec$^{-1}$ for $\varepsilon$ and 66.7mM/sec for $G_{glia}$ is used throughout this paper, where $G_{glia}$ is the glial buffering strength.

To complete the description of the $[K]$ dynamics, we make the assumption that the flow of $Na^+$ into the cell is compensated by flow of $K^+$ out of the. Then the internal [K] ($[K]_i$) can be approximated by

$$[K]_i^{e/i} = 140.0mM + (18.0mM - [Na]_i^{e/i}), \quad (8)$$

where 140.0mM and 18.0mM respectively are the normal resting concentrations of $K^+$ and $Na^+$ inside the cell.

The internal and external $Na^+$ concentrations ($[Na]_i$, $[Na]_o$) are also updated in the model as,

$$\frac{d[Na]_i^{e/i}}{dt} = 0.33\frac{I_{Na}^{e/i}}{\beta} - 3I_{pump}^{e/i}$$

$$[Na]_o^{e/i} = 144.0mM - \beta([Na]_i^{e/i} - 18.0mM), \quad (9)$$

where 144.0mM in equation 9 is the normal resting extracellular $Na^+$ concentration. Further details on $I_{pump}$, $I_{diff}$, $\varepsilon$, $G_{glia}$, $[K]_i$, $[Na]_i$, and $[Na]_o$ can be found in Cressman, et at. (Submitted).

The reversal potential for leak current used in equation 2 is updated based on instantaneous ion concentrations inside and outside the cell using the Goldman-Hodgkin-Katz equation

$$V_L^{e/i} = 26.64 \ln\left( \frac{[K]_o^{e/i} + 0.065[Na]_o^{e/i} + 0.6[Cl]_i^{e/i}}{[K]_i^{e/i} + 0.065[Na]_i^{e/i} + 0.6[Cl]_o^{e/i}} \right), \tag{10}$$

where $[Cl]_i^{e/i}$ and $[Cl]_o^{e/i}$ are the concentrations of intra- and extracellular chloride and are equal to 6.0mM and 130mM respectively.

*Numerical integration:*

We integrated the model numerically using the fourth-order Runge-Kutta method with a time step of 0.01ms. The spatial diffusion of $[K]_o$ was solved by the forward-difference method with a spatial grid $\Delta x = 10.0\mu m$; note also that the $K^+$ surrounding one cell (e.g. PC) can diffuse to the nearest neighbors in the same layer as well as the next layer (IN). Our results were robust to moderate changes in the ratio and number of PCs and INs. Code for our model is available from ModelDB.[1]

## RESULTS

*Persistent states:*

To study instability in neuronal networks, we investigated a two-layer composite of one-dimensional networks consisting of 100 PCs and 100 INs (see Fig.1). Both neuron types were modeled as single-compartment conductance-based cells using the Hodgkin-Huxley formalism (Gutkin et al., 2001). The spike-frequency adapting current was only included in the PCs, a common property of these cells (Mason and Larkman, 1990; Wang, 1998). In the network, both cell types were connected among themselves (PC-PC and IN-IN) and to each other (PC-IN) through spatially-dependent synaptic profiles,

[1] http://senselab.med.yale.edu/modeldb/

where the average PC-PC connections are smaller than the others (IN-IN, PC-IN). Importantly, our model incorporates dynamic intra- and extracellular ion concentrations, along with the effects of pumps, glia, and diffusion. The details of the network connectivity and ion concentration dynamics are described in the methods section.

We followed the protocol of Gutkin et al. (2001) to induce persistent activity in the network by applying a spatially localized Gaussian excitatory stimulus at time t = 112ms for 20ms. In response, the neurons exhibit discharge spikes that activate the connected neurons, which in turn activate other neurons. For an appropriate choice of synaptic parameters, this positive feedback leads to a self-sustained and spatially-localized activity which outlasts the input stimulus, as shown in Fig. 2a. As in Fig. 3 of Gutkin et al. (2001), we find that this occurs when there is an appropriate balance between excitation and inhibition.

To clarify this notion, we examined the consequences of varying the overall strength of the PC-PC and the PC-IN synaptic strengths. Fig. 2b shows plots of the synaptic parameters $\alpha_{ee}$ versus $\alpha_{ei}$ for three different values of the bath potassium concentration $k_{o,\infty}$ . The region between the dark curves corresponds to parameter sets for which self-sustained and spatially-contained network activity can be observed. With too much excitation (i.e., in the region above the top curve), the activity spreads throughout the network. With too much inhibition (i.e., below the bottom curve), the network shows only transient activity in response to the stimulus. In the region between the curves, the excitation and inhibition are sufficiently balanced such that the stimulus gives rise to activity that is persistent and spatially restricted, such as that shown in Fig. 2a.

We observed that the stability region shifts towards smaller values of $\alpha_{ee}$ when $k_{o,\infty}$ is increased (note the different vertical scales in the three panels of Fig. 2b). Thus, the increase in $[K]_o$ due to $K^+$ diffusion from the bath must be counterbalanced by a decrease in excitatory synaptic strength to maintain stability. Note also that the width of the stability region decreases with increasing $k_{o,\infty}$; we observed a 25 percent decrease in width as $k_{o,\infty}$ increased from 3 to 3.5mM (Fig.2b). These results are consistent with Figs. 3 and 8 of Oberheim et al. (2008), in which it was observed that in the epileptic brain, the density of spines and the volume of dendrites increases, which causes an increase in excitation. Meanwhile, the overlap of astrocytes and increase in their volume enhances $[K]_o$ accumulation due to reduced extracellular space. This rearrangement reduces the glial network's capacity to buffer extracellular $K^+$ and contributes to increased glutamate release (Oberheim et al., 2008). The regions above the upper curves in Fig.2b correspond to such a state.

We also investigated the stability of persistent activity by applying perturbing stimuli. Following Gutkin et al. (2001), we observed that persistent activity can be "turned off" by a strong excitatory stimulus (see Fig. 2a at t=500ms). We then applied more moderate stimuli to persistent activity generated with networks within the stable regions of Fig. 2b. The thin lines inside each stability region demarcate the upper bound of the region that is stable to moderate perturbations. More precisely, if the network is between the lower bound of the region and the thin line, then the network is stable to moderate perturbations (see also the discussion of Fig.3c below). Analogous to the experimental results described above, our model shows that the network is less resistant to epileptic drive when already in a relatively excited state (contrast the three panels of Fig.3b).

Persistent activity can only be initiated using a limited range of input current values (Fig.2c). If $I_{amp}$ is too small, the combined effects of the stimulus and excitatory synaptic activity is not sufficient to enter a maintained persistent state. Excessively high input current, $I_{amp}$ is also ineffective at producing long-lasting behavior, as it shuts down the network by synchronizing the activity of all the neurons (e.g. Fig.2a). There is also an intermediate range of $I_{amp}$ values where the network activity spreads over all of the neurons. So in order of increasing stimulus strength, one can see the following four behaviors in the network (results not illustrated): (1) Transient behavior where the activity decays with time (below the lower range of the regions in Fig.2c), (2) a stable focus of activity that lasts for several seconds (within the regions illustrated in Fig.2c), (3) a state where the activity spreads throughout the network (above the upper range of the regions in Fig.2c), and (4) no activity region where the current synchronizes and stops the activity of all neurons in the network (further to the right of the activity regions shown in Fig.2c). We note that the experimental work of Pinto et al. (2005, Fig.2 of this reference) is also consistent with these observations; in that study, a stimulus threshold for wave initiation and its dependence on inhibition and excitation was described.

*Transition from persistent states to seizures:*

To examine the stability of persistent activity against perturbations, we plot the activity of the network for various values of $\alpha_{ee}$ using the same $k_{o,\infty}$,=3.0 mM. We estimate the activity of the network by counting the number of spikes in the excitatory (PC) population that occur within 50msec non-overlapping time windows, and calculate the average (spikes/msec). We take the value of $\alpha_{ee}$ such that it either lies at the boundary or within the region where we see a stable focus of activity (shown in Fig.2b, left panel). An initial excitatory stimulus (at t = 112ms for 20ms) causes a stable focus of activity to appear. After some time, we apply a $2^{nd}$ stimulus (at t = 500ms for 20ms) to all neurons in the network. The second stimulus is stronger than the first so that it can recruit all neurons in the network. As shown in Fig.3a, we observe different behaviors depending on the value of $\alpha_{ee}$: for values within (0.215-0.216), the network returns to the spatially restricted persistent activity that was present before the perturbation. In Fig.3b we show an example of this behavior through a raster plot. However, for larger $\alpha_{ee}$ (0.217), the network fails to regain its stable focus of activity and all the cells in the network persistently fire action potentials. Inhibition dominates at smaller $\alpha_{ee}$ (0.214) and the excitatory perturbation causes the network to decay to silence. The stability of a network persistent state depends on the balance of excitatory and inhibitory synaptic strength of the network. Consistent with our findings, Pinto et al., (2005) found that the permeability of cortical slices to epileptic wave initiation and spreading decreased by increasing 6,7-Dinitroquinoxaline-2, 3-dione (AMPA receptor antagonist). An opposite effect was observed by raising the picrotoxin ($GABA_A$ receptor antagonist) concentration (see for example Figs.5, 6 of Pinto et al., 2005).

As mentioned above, stability can be maintained after increasing $k_{o,\infty}$ by reducing the excitability in the network. Increasing $k_{o,\infty}$ leads to increases in the local potassium concentration $[K]_o$. In order to investigate how this effects stability, we repeated the simulations in Fig.3a using three different values of $k_{o,\infty}$ (different $[K]_o$ values can also be achieved by varying the glial buffering strength $G_{glia}$, however, the results of these simulations remain unchanged). The values of $\alpha_{ee}$ are chosen such that they are always at the center of the three stability regions shown in Fig.2b. It is clear from Fig.3c that for low $k_{o,\infty}$ values (3.0-3.5mM), the activity of the network drops back to the level observed prior to the perturbation. However, the network is unable to regain its stable focus of activity after perturbation when $k_{o,\infty}$ is higher (4.0mM). As mentioned earlier, the region of stability shrinks with increased $[K]_o$. At the same time the percentage of the activity region that is stable to perturbation also decreases with increasing $k_{o,\infty}$. In other words, for about 80 percent of the activity region for $k_{o,\infty}$ = 3.0mM in Fig.2b (from lower bound to the thin solid line with bullets), the network activity is stable to perturbations of moderate size. When $k_{o,\infty}$ = 3.5mM and 4.0mM, the network is stable only for about 50 and 30 percent (between thin solid line with bullets and lower bound) of the persistent activity regions respectively (see Fig.2b). We note that in experimental work, an accelerated spread of epilepsy has been observed by elevating $K^+$ in the bath solution (Figs.2, 3 of Konnerth et al., 1984).

Recently, several groups have reported that the $Ca^{2+}$ dependent glutamate release from glial cells generates slow transient inward currents in nearby neurons through the activation of NMDA receptors (Parpura et al., 1994; Parpura and Haydon, 2000). These discoveries of neuron-glia interaction have been extended by Tian et al., (2005) and Kang et al. (2005) to glial contributions to epilepsy. They showed that glial cells appear to be capable of synchronizing the activity of adjacent neurons through simultaneous non-synaptic slow inward neuronal currents. Another group has made a contrasting claim that glutamate released by glia is not necessary for the generation of epileptic activity in hippocampal slices (Fellin et al., 2006). We use our computational model to differentiate the conditions under which the glial-based perturbations would cause the neuronal network to go to a seizure-like state from that where the glial perturbations are insufficient for the network to make the transition from normal to seizure-like state.

In the framework of our model, the net effect of glutamate released by glial cells will manifest itself as an increased excitatory synaptic drive ($\alpha_{ee}$) in the network. To test the robustness of the network against glial induced excitability, we began our simulation with normal network activity as previously described. After establishing a stable focus of activity we increased the excitability of the network by increasing $\alpha_{ee}$ above the region of stable activity for 30ms (i.e. we applied a second stimulus in $\alpha_{ee}$). The abrupt increase in $\alpha_{ee}$ thus mimics the effect of sudden glutamate release from glia. The activity of the network spread throughout the network but whether or not the persistent state could be reestablished depended on the baseline value of $\alpha_{ee}$. The network is stable to this perturbation for baseline values of $\alpha_{ee}$ = 0.215-0.216 (Fig.4). However, the rate at which the network restored its stable focus of activity depends on the baseline $\alpha_{ee}$ value. If the baseline value of $\alpha_{ee}$ is too small ($\leq$ 0.214), the transient $\alpha_{ee}$ increase destroys the persistent state. Similarly, for sufficiently large $\alpha_{ee}$ values the network was unable to restore its stable focus of activity and the analogue of epileptiform spread following perturbation is shown in the upper trace of Fig.4 (analogous to Figs.4 and 5 of Tian et al. (2005), where they observed epileptiform activity caused by increased excitation due to $Ca^{2+}$ dependent glutamate release by glia). This behavior also depends on the extracellular $K^+$ concentration. For higher $[K]_o$ ($[K]_o \geq$ 4.0mM) values the network cannot reestablish its stable activity when the excitability is increased and then restored (result not shown).

*Spontaneous seizures in the network:*

Next we investigated the dynamics of the network as $[K]_o$ evolves in time. Since the $K^+$ dynamics is very slow we simulated the network for longer times. For the simulation in Fig.5a-c, $k_{o,\infty}$ is taken equal to 14.0mM, which generates very slow oscillations in $[K]_o$ (see Cressman et al., Submitted). In Fig.5a we show the membrane potential for one PC in the network. Initially the cell is silent when $[K]_o$ is low (Fig.5b). However, as $[K]_o$ rises, the cell starts seizure-like tonic firing. The firing frequency increases as $[K]_o$ increases, followed by depolarization block. The spiking returns when $[K]_o$ starts decreasing. The cell eventually goes to a silent state after $[K]_o$ has fallen below a certain value and the whole process repeats for the next cycle of $[K]_o$. The locking of the cells into depolarization block has been observed in experiments (Fig.2 of Bikson et al., 2003; Fig.2 of Ziburkus et al., 2006) and in simulation (Fig.3 of Kager et al., 2007). This effect has been shown to quench synaptic transmission and could underlie the interplay seen between PC and IN during in vitro seizures if inhibition is differentially lost through depolarization block (Fig.2 of Ziburkus et al., 2006). In Fig.5b we show the traces for $[K]_o$ and $[Na]_i$, while the excitatory activity of the network is shown in Fig.5c. The activity drops to zero at higher $[K]_o$ values when the whole network transiently goes into depolarization block. In this slow time-scale setting the intervals between seizures are similar to that seen in seizure experiments (Fig.2 of Ziburkus et al., 2006). Long-time scale intervals with network refractoriness have been also observed in cortical oscillations (Huang, et al., 2004), neonatal burst firing (Leinekugel, et al., 2002), spinal cord burst firing (Chub et al., 2006), up states (Shu, et al., 2003), and interestingly in spreading depression (Somjen, 2004; Kager, et al., 2000).

After enhancing the glial uptake slightly (compared to the simulation in Fig.5a-c) depolarization block disappears and all the neurons in the network show high frequency, lower amplitude seizure-like spiking when $[K]_o$ reaches a sufficiently high value during $[K]_o$ oscillations (Fig.5d-e). Like the previous example, each cell starts with slow spiking initially as $[K]_o$ starts rising (Fig.5dA) followed by rapid tonic firing at high $[K]_o$ (Fig.5dB). Finally, the network switches to a silent state as $[K]_o$ slowly falls to lower values (end of Fig.5dC) until the next cycle. In Fig.5e, we show $[K]_o$ and $[Na]_i$ for the same PC shown in Fig.5d. The increases in $[K]_o$ during seizures have been observed in other studies (Bazhenov et al., 2004; Fig.1 of Bikson et al., 2003; Frohlich et al., 2008; Fig.8 of Kager et al., 2000; Kager et al., 2007). The activity diagram (Fig.5f) shows a clear peak at high $[K]_o$ values indicating increased but still partial (notice that the activity does not rise to 100 during seizures) spiking synchrony in the network. However, the longer time-scale oscillations (seizure episodes and $[K]_o$ oscillations) are more synchronized through the entire network. In Fig.5dD we show a raster plot for the network at high $[K]_o$ value during the cycle described above. The network shows a stable focus of activity for smaller $[K]_o$ values, which changes into seizure-like behavior for larger $[K]_o$ values (Fig.5f), and interstingly is asynchronous, consistent with experimental (Figs.11 and 12 of Netoff and Schiff, 2002) and theoretical (Fig.2 of Gutkin, et al., 2001) findings.

## DISCUSSION

We have used a one-dimensional two-layer network model to study the physiological conditions under which a neuronal network can switch to a persistent state of activity, and the stability of the persistent state to perturbations. Previously, several experimental and theoretical studies have shown that spatially localized persistent activity in neuronal networks requires a balance between excitatory and inhibitory synaptic inputs (Compte et al., 2000; Funahashi et al., 1989; Gutkin et al., 2001). We show that the network not only needs balanced excitatory and inhibitory synaptic footprints, but also balanced levels of extracellular $K^+$, controlled by the glial syncitium, to achieve local persistent activity that is stable to perturbations.

Our study successfully models persistent activity during both the up state (Sanchez-Vives and McCormick, 2000), and during the delay periods of working memory tasks using recurrent excitation balanced by local inhibition (Compte et al., 2000; Gutkin et al., 2001; Wang, 1999). The persistent activity in our model can last for several seconds which has been observed in several experimental preparations (Fuster, 1995; McCormick et al., 2003). In addition, the firing rate of our model PCs during the persistent activity is similar to that seen during the delay responses in memory tasks in vivo (Funahashi et al., 1989; Miller et al., 1996).

In recent years, neuroscience research has unveiled a previously unimaginable complexity of glial cells. Astrocytes, in particular have been shown to affect virtually every aspect of neuronal function through glia-neuron cross talk. They have the ability to buffer excitatory ions such as $K^+$ from the interstitial volume (Amzica et al., 2002), preventing an undue increase in excitation of the neuronal network. They can also enhance the excitability of the network by releasing glutamate and ATP in a $Ca^{2+}$ dependent manner, which can activate NMDA and AMPA receptors synaptically and non-synaptically (Parpura et al., 1994; Parpura and Haydon, 2000). In spite of a direct role of glia in the activity modulation of neuronal networks, detailed models of glia-neuron interaction are few (Nadkarni and Jung, 2003). Here we use a model to show that dysfunction of glial cells can cause the transition from normal stable neuronal activity to seizure-like uncontrolled activity. Several experimental studies support glial dysfunction during epileptic seizures (Heinemann et al., 2000; Hinterkeuser et al., 2000).

The ability of astrocytes to release glutamate in a $Ca^{2+}$ dependent manner has been recently extended by two groups to the glial bases of epilepsy (Tian et al., 2005; Kang et al., 2005). We tested the stability of the network to glial-induced excitatory perturbations by perturbing the network from its stable focus of activity through briefly raising the excitatory weight of the network. Our simulation shows that the ability of glial cells to induce seizure-like activity depends on the initial baseline state of the network. If the network is in a relatively low excitatory state, then a transient glial glutamate perturbation would not be able to cause epileptic activity no matter how strong the perturbations are. However, even a small perturbation by glial cells is enough to cause seizure-like activity in the network if its excitability set point is sufficiently high.

Recent experimental results point to the role of astrocytic domain reorganization in the epileptic brain (Oberheim, et al., 2008). To test the effect of realistic cells' layout and spatial arrangement of glia, spatially explicit models are needed. Such models would have to take into account the effect of extracellular tortuosity and explicit molecular diffusion between network elements, something beyond the scope of the present study.

There is an extensive and deep literature that describes that the stability of ensemble systems (Forster, 1990), and indeed the fundamental hallmark of a stable (equilibrium) system in physics is reflected in how such systems handle fluctuations. We hypothesize that an analogous principle regarding the response to perturbations applies to characterize the stability of biological systems. We therefore examined the response of our extended network of INs and PCs to fluctuations. First we delivered a perturbation, which created a persistently active state and then delivered a second perturbation. Depending upon the excitability of the network, which we could set through the synaptic efficacy (Fig.3a), or the extracellular $K^+$ level (Fig.3c), the response of the persistent state to the second perturbation might be exponential decay back to the persistent state, a runaway expansion of activity to break through the inhibitory constraints and involve more of or the whole network, or destruction of the persistent state with quiescence (behaviors illustrated in Fig.3a and 3c). Such competition between excitation and inhibition in response to perturbation appears to be reflected in the recent experimental results in an 'inhibitory veto' in such networks in cortex (Trevalyan et al., 2006).

Our findings are an extension of previous model studies that demonstrated bistable states in networks, i.e., to fulfill the working memory function, these models have found the physiological conditions under which the network can switch back and forth between a resting state and spatially structured activity induced by transient inputs (see for example Compte et al., 2000). We extended these findings to identify the conditions under which a network displaying a stable persistent activity can switch to seizure-like states. The maintenance of a low $[K]_o$ is crucial to create persistent activity robust to perturbations.

Neurophysiologic studies of working memory by Miller et al. (1996) have shown that the delay activity in the prefrontal cortex of monkeys is resistant to occasional distractions. Through mathematical modeling these results were confirmed by Compte et al. (2000). This latter study related the distraction-resistant activity to the stimulus strength; that is, the activity packet was predicted to be resistant to small perturbations. The main finding of our study is that the network activity packet is stable provided that (1) the excitable synaptic strength is not too high (a parameter modulated by glial glutamate release); (2) $[K]_o$ is low enough to be within the physiological range (i.e. the glial network buffer is functioning properly); and (3) the perturbations are not too strong. Thus our study asserts two additional conditions on the stability of the neuronal activity. Even small perturbations can push the network to an unstable state if the network is already in a state of increased $\alpha_{ee}$ or the $[K]_o$ is relatively high. Both of these conditions are regulated by a properly functioning glial network.

The evolution of activity seen in Fig.5a and 5d is intimately linked to the ionic dynamics shown in Fig.5b and 5e. Without incorporating such ionic and glial dynamics, such evolution would not appear. Seizure patterns, whether observed in vitro (Ziburkus et al., 2006), or from human patients recorded from scalp or intracranial electrodes (Schiff et al., 2005), demonstrate a prominent and very consistent evolution from initiation to termination. Whether such evolution observed from such different scales of experimental observation is related to the ionic and glial dynamics studied here is an intriguing question that warrants further exploration.

Although brains are not equilibrium systems, we propose that how neuronal networks respond to perturbations, and how the response to such perturbations decays (or grows), are critical determinants of the state of the brain, with parallels to other physical systems (Forster, 1990). How neuronal networks respond to perturbation determines how transient patterns of activity emerge from a background activity in response to internal fluctuations of activity or external stimuli. We propose that the response of a balanced, or mildly imbalanced network, to such perturbations is a fundamental feature which may underlie a variety of disparate transient phenomena such as seizures, working memory, up states, cortical oscillations, and spinal cord and hippocampal neonatal burst firing.

**ACKNOWLEGEMENTS**

We thank Jokubas Ziburkus, Andrew J Trevelyan, Maxim Bazhenov, and Partha Mitra, for their valuable discussions. This work was funded by NIH Grants K02MH01493 (SJS), R01MH50006 (SJS, GU), F32NS051072 (JRC), and CRCNS-R01MH079502 (EB).

**FIGURE LEGENDS**

**Fig. 1:** Topology of the network. The network consists of one pyramidal cell layer and one interneuron layer, both arranged in a ring. The two neuronal types make synaptic connections in the same layer as well as across the layers with Gaussian synaptic footprints. The $K^+$ concentration around each neuron diffuses to the nearest neighbors in the same layer and the nearest neighbor in the next layer (arrows).

**Fig. 2:** The existence and stability of spatially restricted foci of activity in our neuronal network. (a) A 20ms excitatory Gaussian stimulus with amplitude $I_{amp}$ = 1.5μA/cm$^2$ (see text) applied to PCs 21-79 causes a persistent and spatially restricted activity packet in the network. This activity is turned off by synchronizing neuronal activity using a strong 1ms stimulus of 100μA/cm$^2$ at $t$ = 500ms. The top (bottom) panel shows a raster plot for the PC (IN) network where each dot represents a single-cell spike in the network. Here $\alpha_{ee}$ = 0.215, and $k_{o,\infty}$ = 3.0mM. (b) The region for persistent and spatially restricted focus of activity using three different $k_{o,\infty}$ values. The focus is persistent, spatially restricted and is turned off by a strong synchronizing stimulus in the region enclosed by two lines (thick solid lines with open circles for three $k_{o,\infty}$ values). Below this region the inhibition is stronger than excitation and the focus does not exist, while above this region the excitation is stronger as compared to inhibition and the activity spreads throughout the entire network. The $\alpha_{ee}$ values for a stable activity focus decreases as $k_{o,\infty}$ is increased (2b, middle and right panels) showing that for persistent states, the increase in $k_{o,\infty}$, must be balanced by a decrease in excitation (note the different ordinate scales). The thin solid lines with bullets inside all activity regions represent the upper bound of the activity regions that are stable against perturbations of fixed (moderate) size. i.e. if the network is between the lower bound of the activity region and the thin line with bullets, then the network is stable against moderate perturbations. Above the thin line in each region the network can show persistent and spatially restricted foci of activity but is not stable against moderate perturbations (although it is still stable against small perturbations). It is clear that the percentage of the activity region that is stable to perturbations decreases with increasing $k_{o,\infty}$. The existence of the focus also depends on the stimulus strength.

The focus does not exist if the stimulus is too weak (c). After a certain threshold of the stimulus ($I_{amp}$ = 0.31 here) the weakness of the stimulus is compensated by the excitation strength and external potassium level. A value of $\alpha_{ei}$ = 0.2 is used for simulations in (a, c).

**Fig. 3:** The activity (see text for definition) of the excitatory population for various $\alpha_{ee}$ values (a). 1$^{st}$ stimulus of Gaussian shape (see text) at t = 112ms for 20ms causes a stable focus of activity in the network. A 2$^{nd}$ stimulus of 1.0μA/cm$^2$ at t = 500ms for 20ms causes the focus to take over the entire population. After the stimulus is removed the network either returns to the stable focus or remains in the state where the activity is spread throughout the entire population depending on the excitation strength. The numbers on the right of each line are the $\alpha_{ee}$ values used during the simulations. In (b) we show a raster plot for a single simulation in (a). $\alpha_{ee}$ = 0.214 is used in this example. (c) Same as (a) but here we use different $k_{o,\infty}$ values given in the plot. The stability of the network also depends on the steady state $[K]_o$. For smaller extracellular $K^+$ the network activity decays back to the stable focus after the 2$^{nd}$ stimulus is turned off. However, if $[K]_o$ is high the network maintains its enhanced activity even after the 2$^{nd}$ stimulus vanishes. $\alpha_{ee}$ values are chosen such that they lie at the centers of the activity regions shown in Fig.2b (right column in Fig. 3c). For all simulations in (a-c) $\alpha_{ei}$ = 0.2 is used. The curves in panels (a and c) are averaged over several trials.

**Fig. 4:** Same as Fig.3a, c but the 2$^{nd}$ stimulus is in $\alpha_{ee}$, i.e. instead of applying current for the 2$^{nd}$ stimulus the $\alpha_{ee}$ is increased beyond the stable activity region (see Fig.2b) to a value of 0.25 for 30ms at $t$ = 500ms. After the $\alpha_{ee}$ value is restored the network returns to the stable focus of activity if the baseline $\alpha_{ee}$ is in an intermediate range. If the baseline $\alpha_{ee}$ is too large the network cannot return to the stable focus. Similarly if the baseline $\alpha_{ee}$ is too small the persistent state decays after the perturbation is removed. The baseline $\alpha_{ee}$ values are given in the plot and $\alpha_{ei}$ = 0.2.

**Fig. 5:** Spontaneous activity in the network for varying $[K]_o$. In panel (a) we show the membrane potential of one PC in the network as $[K]_o$ (solid line) and $[Na]_i$ (dashed line) evolve in time (b). (c) shows the activity of the network, where the drop in activity occurs at high $[K]_o$ values due to the switching of the network to depolarization block. For (a-c) $G_{glia}$=16.7 *and* $\varepsilon$ =0.23. In (d-f) we repeat the simulations in (a-c) using $G_{glia}$=26.7 *and* $\varepsilon$ =0.003. For these parameters all cells in the network show high frequency seizure-like firing for higher $[K]_o$ values instead of going into depolarization block. As clear from (dA) each cell exhibits regular spiking initially but shifts to seizure-like bursting for higher $[K]_o$ values (dB), and finally to a silent state (end of dC). There is a clear peak in the activity of the network at high $[K]_o$ values, indicating the increased synchrony in the network (f). The raster plot at peak activity is shown in (dD). The solid and dashed lines in (e) show $[K]_o$ and $[Na]_i$ trajectories respectively for the PC shown in (d). Parameter values used for simulation in this figure are $k_{o,\infty}$=14mM, $\alpha_{ee}$=0.12, *and* $\alpha_{ei}$=0.2.

Table 1: Model variables and parameters.

| Variable | Description | Variable | Description |
| --- | --- | --- | --- |
| $V$ | Membrane potential | $[K]_o$ | Extracellular potassium concentration |
| $I_K$ | Potassium current | $[K]_i$ | Intracellular potassium concentration |
| $I_{Na}$ | Sodium current | $g^{ij}$ | Spatial extent of synaptic current from ith to jth neuron |
| $I_L$ | Leak current | $\chi$ | Variable modeling the depolarization block |
| $I_{syn}$ | Synaptic current | $\eta$ | Threshold to synaptic block due to the depolarization |
| $I_{ext}$ | Stimulus current | $[Na]_o$ | Extracellular sodium concentration |
| $I_{rand}$ | Random background noise | $[Na]_I$ | Intracellular sodium concentration |
| $s$ | Variable giving the temporal evolution of synapse | | |
| $V_L$ | Reversal potential of leak current | | |
| $I_{pump}$ | Pump current | | |
| $I_{glia}$ | Glial uptake | | |
| **Parameter** | **Description** | **Parameter** | **Description** |
| $C$ | Membrane capacitance | $\gamma$ | Forward transitions into and out of synaptic block due to the depolarization |
| $I_{amp}$ | Stimulus amplitude | $\tilde{\gamma}$ | Backward transitions into and out of synaptic block due to the depolarization |
| $\tau$ | Time Constant of synapses | $D$ | Diffusion coefficient of potassium |
| $g_L$ | Conductance of leak current | $[Cl]_o$ | Extracellular chloride concentration |
| $g_{AHP}$ | Conductance of afterhyperpolarization current | $[Cl]_i$ | Intracellular chloride concentration |
| $k_{o,\infty}$ | Steady state extracellular concentration | | |
| $G_{glia}$ | Strength of glial uptake | | |
| $\alpha_{ij}$ | Strength of synaptic current from ith to jth neuron | | |
| $\varepsilon$ | Constant for potassium diffusion to bath solution | | |

**FIGURES**

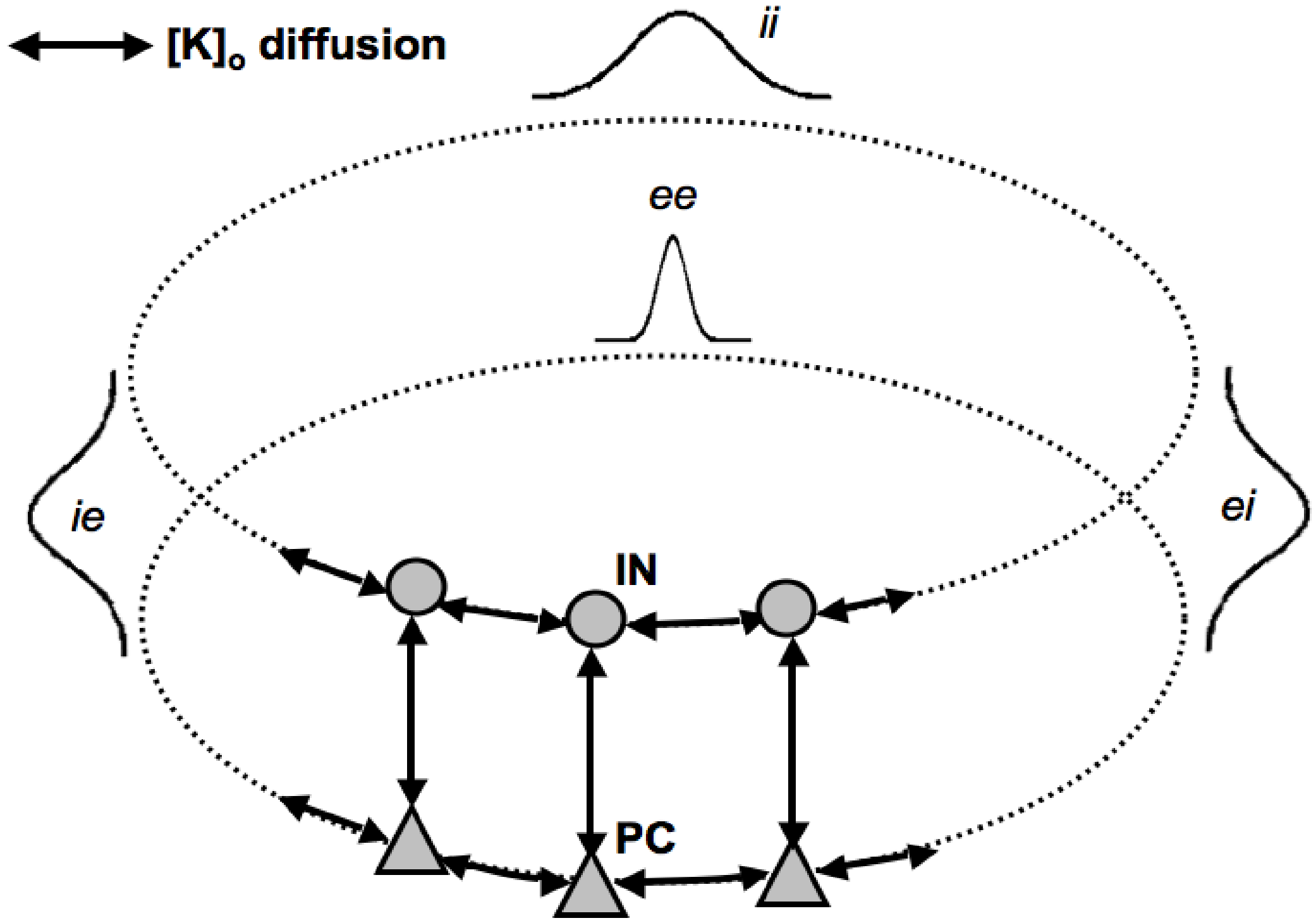

Fig. 1.

**FIGURES**

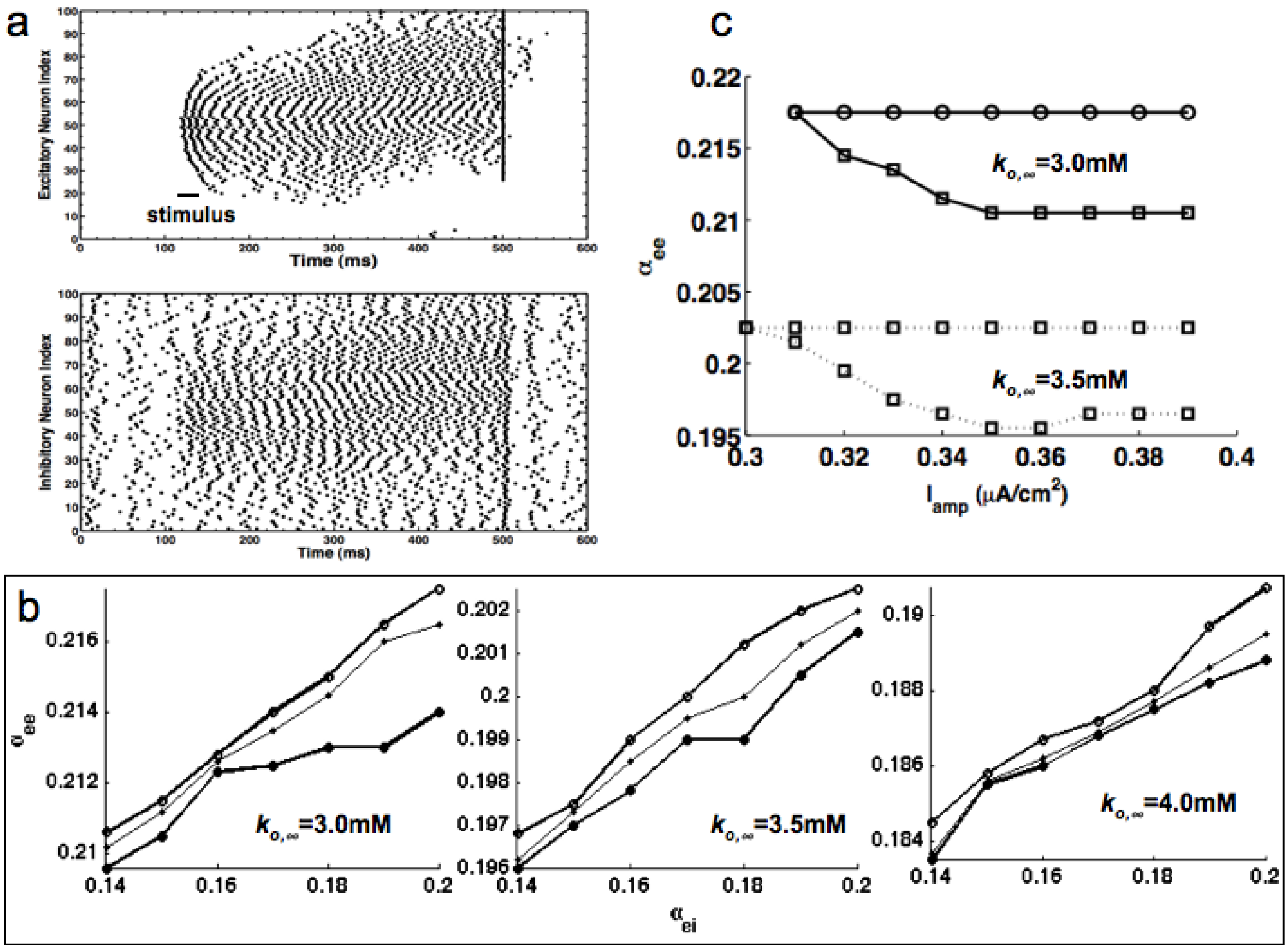

Fig. 2.

**FIGURES**

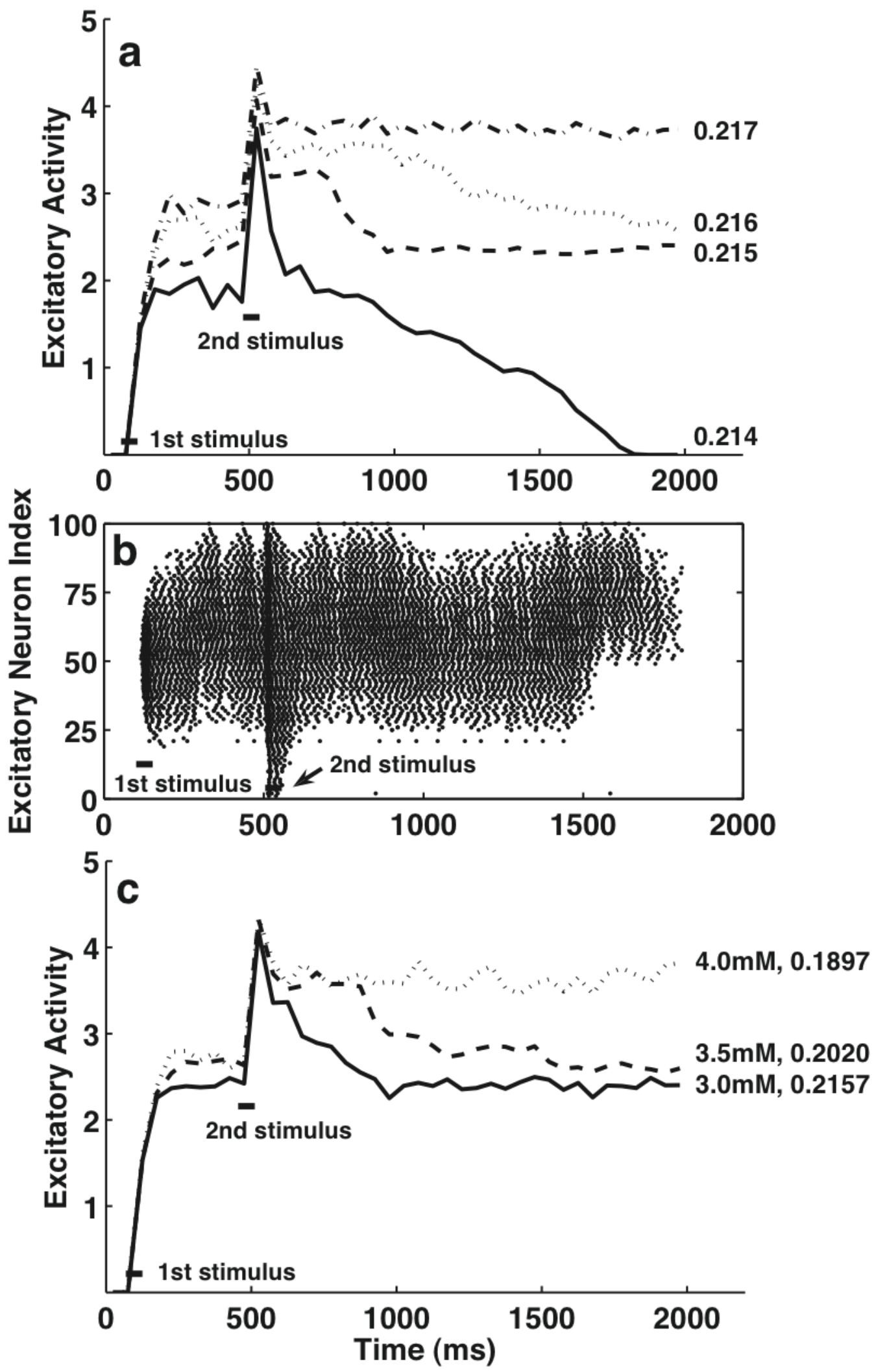

Fig. 3.

**FIGURES**

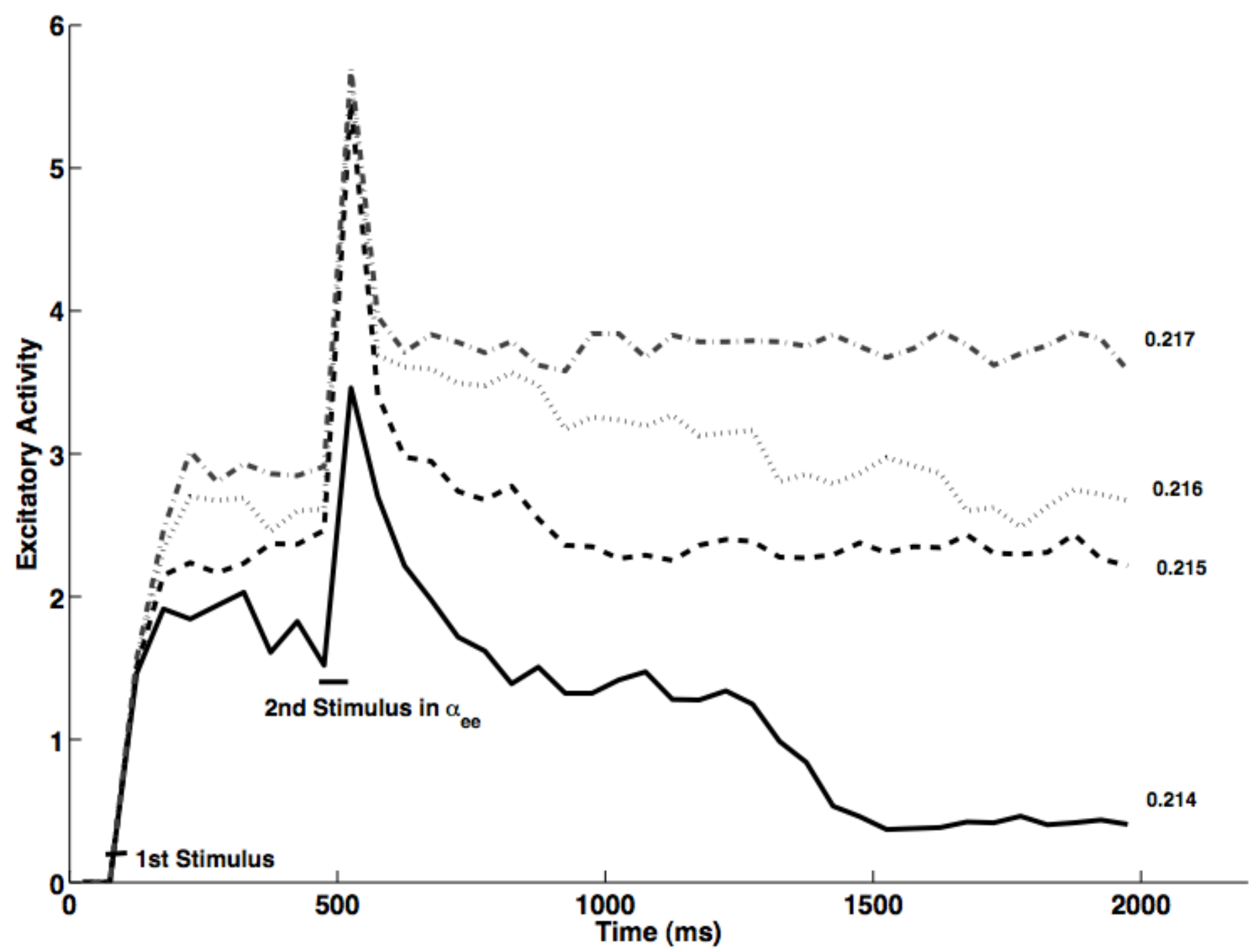

Fig. 4.

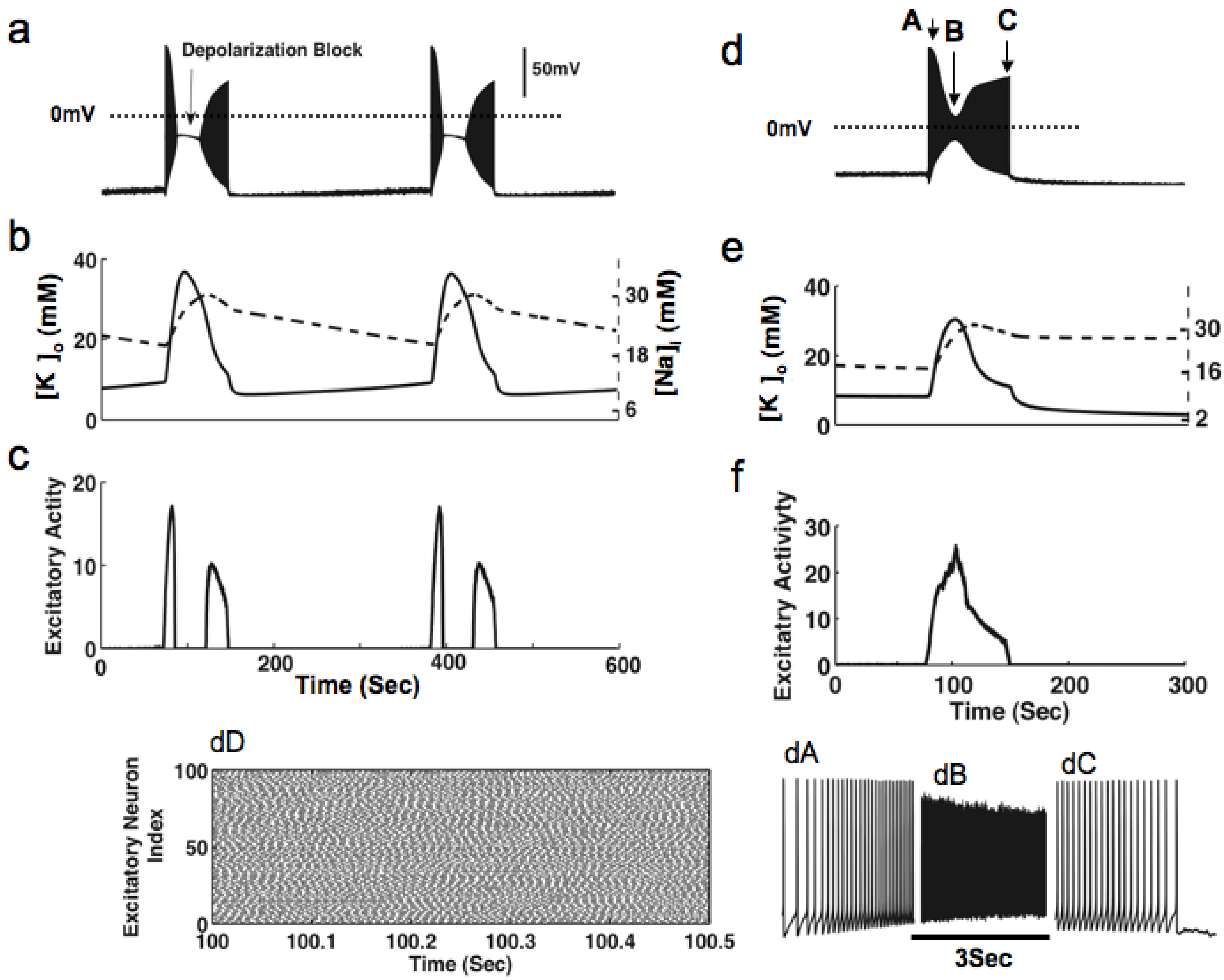
a
Depolarization Block
50mV
0mV
b
[K ]o (mM)
[Na]i (mM)
40
20
0
30
18
6
c
Excitatory Actity
20
10
0
0
200
400
600
Time (Sec)
dD
Excitatory Neuron Index
100
50
0
100
100.1
100.2
100.3
100.4
100.5
Time (Sec)
d
A
B
C
0mV
e
[K ]o (mM)
[Na]i (mM)
40
20
0
30
16
2
f
Excitatry Activiyty
30
20
10
0
0
100
200
300
Time (Sec)
dA
dB
dC
3Sec

Fig. 5